\documentclass[11pt,a4paper]{article}

\usepackage{amsmath}
\usepackage{amssymb}
\usepackage{layaureo}
\usepackage[utf8]{inputenc}
\usepackage{url}
\usepackage{hyperref}
\usepackage{graphicx}

\title{Sub-horizon evolution of cold dark matter perturbations through dark matter-dark energy equivalence epoch}
\author{O.~F.~Piattella\footnote{oliver.piattella@pq.cnpq.br}, D.~L.~A.~Martins\footnote{denilsonluizm@gmail.com}~~and L.~Casarini\footnote{casarini.astro@gmail.com}\\
Department of Physics, Universidade Federal do Esp\'irito Santo,\\
avenida Ferrari 514, 29075-910 Vit\'oria, Esp\'irito Santo, Brazil}
\date{}

\begin{document}

\maketitle

%%%%%%%%%%%%%%%%%%%%%%%%%%%%%%%%%%%%%%%%%%%%%%%%%%%%%%%%%%%%%%%%%%%%%%%%%%%%%%%%%%%%%%%%%%%%%%%%%%%%%%%%%%%%%%%%%%%%%

\begin{abstract}
We consider a cosmological model of the late universe constituted by standard cold dark matter plus a dark energy component with constant equation of state $w$ and constant effective speed of sound. Neglecting fluctuations in the dark energy component we obtain an equation describing the evolution of sub-horizon cold dark matter perturbations through the epoch of dark matter-dark energy equality. We explore its analytic solutions and calculate an exact $w$-dependent correction for the dark matter growth function, logarithmic growth function and growth index parameter through the epoch considered. We test our analytic approximation with the numerical solution and find that the discrepancy is less than 1\% for $k =0$ in the epoch of interest.
\end{abstract}

%%%%%%%%%%%%%%%%%%%%%%%%%%%%%%%%%%%%%%%%%%%%%%%%%%%%%%%%%%%%%%%%%%%%%%%%%%%%%%%%%%%%%%%%%%%%%%%%%%%%%%%%%%%%%%%

\section{Introduction}

Current observations of the cosmic microwave background (CMB), type Ia supernovae, baryon acoustic oscillations (BAO) and large scale structures (LSS) constrain the present dark energy (DE) parameter, $w_0$, around $-1$ (corresponding to a  cosmological constant). For example, from the Seven-Year Wilkinson Microwave Anisotropy Probe (WMAP) analysis \cite{Komatsu:2010fb} one of the best fit values is $w_0 = -1.10\pm 0.14$ (68\% Confidence Level), obtained without taking into account high redshift supernovae. Including the latter moves the best fit value across $-1$: $w_0 = -0.980\pm 0.053$ (68\% CL). From the Nine-Year WMAP analysis \cite{Hinshaw:2012aka} the above constraints are slightly improved, e.g. $w_0 = -1.084 \pm 0.063$ (68\% CL). From the analysis of the Union2 supernovae data set combined with CMB data \cite{Amanullah:2010vv}, $w_0 = -0.997^{+0.077}_{-0.082}$, with combined statistical (68\% CL) and systematic errors. Latest Planck results \cite{Ade:2013zuv} also do not change dramatically this picture, providing $w_0 = -1.49^{+0.65}_{-0.57}$ (95\% CL) for CMB only data (including polarisation) and $w_0 = -1.13^{+0.24}_{-0.25}$ (95\% CL) when also including BAO data. Other best fits, involving different cosmological probes, can be found in the LAMBDA website.\footnote{\url{http://lambda.gsfc.nasa.gov/product/map/dr4/parameters.cfm}}

These results do not exclude the possibility of DE being a dynamical component of our universe, including a phantom \cite{Zlatev:1998tr, Tsujikawa:2013fta, Piattella:2013wpa, Caldwell:2003vq}. Therefore, an interesting issue is to learn how DE dynamics affects the late-time evolution of cold dark matter (CDM) perturbations. This issue recalls M\'{e}sz\'{a}ros equation \cite{Meszaros:1974tb}, which describes the sub-horizon evolution of CDM perturbations in a CDM + radiation scenario. Here, we neglect radiation and take into account DE, tracking the evolution of sub-horizon CDM perturbations through the epoch of DM-DE equality. This idea was put forward for the first time, up to our knowledge, by the authors of \cite{Boehmer:2010hg}, who named the equation found in their paper as \textit{$w$-M\'{e}sz\'{a}ros equation}.
%Unfortunately, this paper seems to contain a mistake in eqs.~(17) and (18) therein. This can be checked, for example, noting that the CDM density contrast has not the standard behaviour (linearly growing with the scale factor) in the DM-dominated limit. Therefore, our first scope in this paper is to find the correct from of the \textit{$w$-M\'{e}sz\'{a}ros equation} (so dubbed by the authors of \cite{Boehmer:2010hg}).

M\'{e}sz\'{a}ros' calculations \cite{Meszaros:1974tb} are performed neglecting the contribution of radiation at the perturbative order. Analytic solutions are given also in \cite{GrothPeebles1975} and \cite{Hu:1995en} (in the latter the baryon component is also taken into account at the background level). On the other hand, a full justification for neglecting radiation perturbations is finally given by Weinberg \cite{Weinberg:2002kg}. Variations or improvements of the calculations in \cite{Boehmer:2010hg} can be found for example in \cite{Vale:2001ng}, where the authors consider the evolution of perturbations in presence of a cosmological constant and adopt both a relativistic as well as a Newtonian description. A thorough mathematical description of the case with CDM plus hot dark matter (HDM) plus baryons is analysed in \cite{Gailis:2006tm, Gailis:2006tn}.

Our paper is organised as follows. In sec.~\ref{sec:2} we present the model and the basic set of equations which describe the expansion of the universe and the evolution of small, linear, fluctuations. In sec.~\ref{sec:3} we consider the small-wavelength limit and derive the $w$-Mészáros equation. In sec.~\ref{sec:4} we present and classify its solutions. In sec.~\ref{sec:5} we consider the asymptotic behaviour of the solutions in order to match them to the well-known solution in the matter-dominated era. In sec.~\ref{sec:gamma} we determine an analytic formula for the DM density contrast, growth function, logarithmic growth function and growth index parameter and analyze its goodness by comparing it with the numerical solution. Finally, in sec.~\ref{sec:6} we discuss our conclusions. Throughout the paper we shall use units $c = 1$.

%%%%%%%%%%%%%%%%%%%%%%%%%%%%%%%%%%%%%%%%%%%%%%%%%%%%%%%%%%%%%%%%%%%%%%%%%%%%%%%%%%%%%%%%%%%%%%%%%%%%%%%%%%%%%%%%

\section{Basic equations}\label{sec:2}

Our background geometry is a flat Friedmann-Lemaître-Robertson-Walker (FLRW) metric:
\begin{equation}
 ds^2 = -dt^2 + a(t)^2\delta_{ij}dx^idx^j\;,
\end{equation}
on which we consider two energy components: one is the usual pressure-less CDM and the other is a DE component described by the equation of state $p_{\rm de} = w\rho_{\rm de}$, with $w$ constant. We also assume the two components not to interact directly, therefore they satisfy separately their own continuity equations, whose solutions are:
\begin{equation}
 \rho_{\rm dm} = \rho_{\rm dm,0}a^{-3}\;, \qquad \rho_{\rm de} = \rho_{\rm de,0}a^{-3(1 + w)}\;,
\end{equation}
where the subscript $0$ refers to a quantity evaluated today and $a_0 = 1$. Employing these solutions, Friedmann equation can be written in the following form:
\begin{equation}\label{FriedmEq}
 \frac{H^2}{H_0^2} = \frac{\Omega_{\rm dm,0}}{a^3} + \frac{\Omega_{\rm de,0}}{a^{3(1 + w)}}\;,
\end{equation}
where $H_0$ is the Hubble constant, $\Omega \equiv 8\pi G\rho/3H_0^2$ is the density parameter and, due to spatial flatness, $\Omega_{\rm de,0} = 1 - \Omega_{\rm dm,0}$. Note that, in order to provide an accelerated expansion we need $w < -(1 + \rho_{\rm dm}/\rho_{\rm de})/3$. As discussed in the Introduction, $w$ has a value about $-1$ today. Note that a perfect fluid model with constant negative $w$ cannot represent DE, unless considering non-adiabatic perturbations, since its square speed of sound would be negative, thereby causing instabilities. The paradigm we have in mind here is a scalar field, possibly non-canonical. See, for example, \cite{Zlatev:1998tr, Tsujikawa:2013fta, Piattella:2013wpa}.

Following \cite{Weinberg:2002kg}, the evolution of perturbations is described by the following system of equations:
\begin{eqnarray}
\label{Eq1} \frac{d}{dt}\left(a^2\psi\right) &=& -4\pi G a^2\left[\rho_{\rm dm}\delta_{\rm dm} + \left(1 + 3c_{\rm de}^2\right)\rho_{\rm de}\delta_{\rm de}\right]\;,\\
\label{Eq2} \dot{\delta}_{\rm dm} &=& - \psi\;,\\ 
\label{Eq3} \dot{\delta}_{\rm de} + 3H\left(c_{\rm de}^2 - w\right)\delta_{\rm de} &=& -\left(1 + w\right)\left(\psi - k^2U_{\rm de}\right)\;,\\
\label{Eq4} \frac{d}{dt}\left[a^5\left(1 + w\right)\rho_{\rm de}U_{\rm de}\right] &=& -a^3c_{\rm de}^2\rho_{\rm de}\delta_{\rm de}\;,
\end{eqnarray}
where $\psi$ is the gravitational potential, $c_{\rm de}^2 = \delta p/\delta\rho$ is the dark energy effective speed of sound, $k$ is the comoving wavenumber and $U_{\rm de}$ is the dark energy velocity potential. The dot denotes derivation with respect to $t$ and the gauge chosen is the synchronous one, with the additional choice of zero CDM velocity, which fixes the residual gauge freedom.

\section{Deep inside the horizon}\label{sec:3}

We derive now the $w$-M\'esz\'aros equation by considering $k/a \gg H$, i.e. perturbations much smaller than the Hubble radius. The question is: can we neglect DE perturbations? Intuitively, if  $c_{\rm de}^2$ is vanishingly small, the answer to that question is no, since DE could cluster. On the other hand, in order to answer the question properly, one should perform an analysis similar to Weinberg's one in \cite{Weinberg:2002kg}, with DE replacing radiation. If $w$ and $c_{\rm de}^2$ are of order of unit, one can show that Weinberg's analysis of slow and fast modes proceed in the same way as in the radiation case. Therefore, being $w$ of order unity (see the introduction), we assume $c_{\rm de}^2$ also of order unity, leaving the case $c_{\rm de}^2 \to 0$ for a future investigation. See for example \cite{Abramo:2007iu, Abramo:2009ne, Dossett:2013npa, Batista:2014uoa} for references where the impact of DE fluctuations on the growth of DM ones is taken into account.

\begin{figure}[h]
\center
\includegraphics[width=0.5\textwidth]{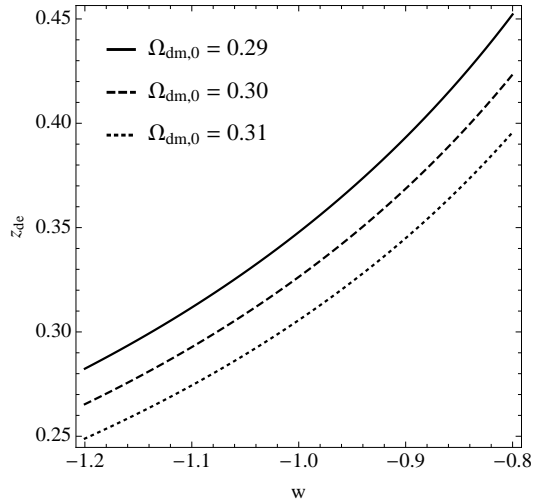}
\caption{Evolution of $z_{\rm de}$ as a function of $w$, from eq. \eqref{OmegaD0aeqrel} with $\Omega_{\rm dm,0}= 0.29$ (solid line), $\Omega_{\rm dm,0}= 0.30$ (dashed line) and $\Omega_{\rm dm,0}= 0.31$ (dotted line).}
\label{figzE}
\end{figure}

Under the conditions above stated, DE perturbations are negligible with respect to DM ones, even when $\rho_{\rm de} > \rho_{\rm dm}$. Thus, eq.~\eqref{Eq1} reads
\begin{equation}\label{wMeszeq}
 \frac{d}{dt}\left(a^2\dot\delta_{\rm dm}\right) = 4\pi G a^2\rho_{\rm dm}\delta_{\rm dm}\;.
\end{equation}
Using the same notation as in \cite{Boehmer:2010hg}, let us define
\begin{equation}\label{ydef}
 y \equiv \frac{\rho_{\rm dm}}{\rho_{\rm de}} = \left(\frac{a}{a_{\rm de}}\right)^{3w}\;,
\end{equation}
where $a_{\rm de}$ is the scale factor at which $\rho_{\rm dm} = \rho_{\rm de}$. It can be related to the present time matter density parameter as follows:
\begin{equation}\label{OmegaD0aeqrel}
\frac{\Omega_{\rm dm,0}}{1 - \Omega_{\rm dm,0}} = a_{\rm de}^{-3w} = (1 + z_{\rm de})^{3w}\;,
\end{equation}
where we used the spatial flatness condition $\Omega_{\rm de,0} = 1 - \Omega_{\rm dm,0}$. In figure \ref{figzE} we plot the evolution of $z_{\rm de}$ as a function of $w$.

Finally, using eq. \eqref{ydef} one can cast eq.~\eqref{wMeszeq} in the following form:
\begin{equation}\label{eqsubhordmnode3}
 \delta''_{\rm dm} + \left[\frac{4 - 3(1 + w)}{6wy} + \frac{2 + 3y}{2y(1 + y)}\right]\delta'_{\rm dm} - \frac{1}{6w^2y(1 + y)}\delta_{\rm dm} = 0\;,
\end{equation}
where the prime denotes derivation with respect to $y$. This is the $w$-M\'esz\'aros equation. Since $\delta_{\rm de}$ is negligible with respect $\delta_{\rm dm}$, the evolution of the latter does not depend on $c_{\rm de}^2$. So, DE interferes with the growth of DM inhomogeneities only via the evolution of the background geometry.

\section{Solutions of w-M\'esz\'aros equation}\label{sec:4}

In the limit $y \gg 1$, i.e. in the matter-dominated phase, eq.~\eqref{eqsubhordmnode3} simplifies to
\begin{equation}\label{eqsubhordmnode3matter}
 \delta''_{\rm dm} + \left(1 + \frac{1}{6w}\right)\frac{1}{y}\delta'_{\rm dm} - \frac{1}{6w^2y^2}\delta_{\rm dm} = 0\;, \qquad (y \gg 1)\;,
\end{equation}
whose general solution is
\begin{equation}\label{eqsubhordmnode3mattersol}
 \delta_{\rm dm} = A\; y^{-1/2w} + B\;y^{1/3w}\;, \qquad (y \gg 1)\;,
\end{equation}
and, with the help of eq.~\eqref{ydef}, one recognises the second mode as the growing one for matter (i.e., $\delta_{\rm dm} \propto a$).
%\footnote{Equation~(19) of \cite{Boehmer:2010hg} does not possess this limit solution, therefore there must be something missing in the calculations there.}

Equation~\eqref{eqsubhordmnode3} can be cast in the form of a gaussian hypergeometric equation \cite{abramowitz1964handbook} 
\begin{equation}
 x(1 - x)\frac{d^2\delta_{\rm dm}}{dx^2} + \left[\gamma - \left(\alpha + \beta + 1\right)x\right]\frac{d\delta_{\rm dm}}{dx} - \alpha\beta \delta_{\rm dm} = 0\;.
\end{equation}
Manipulating eq.~\eqref{eqsubhordmnode3} we can read off the following values for the parameters and the variable:
\begin{equation}
 \alpha = -\frac{1}{3w}\;, \qquad \beta = \frac{1}{2w}\;, \qquad \gamma = \frac{1}{2} + \frac{1}{6w}\;, \qquad x = -y\;.
\end{equation}
Since $y$ is positive definite, we are interested in negative values of $x$ only. Therefore, as it should be, there are no singularities in the evolution of $\delta_{\rm dm}$ since the gaussian hypergeometric function has no singularities for negative real values of its argument. The singularity in $x = y = 0$ corresponds to the remote future $a \to \infty$.

It is possible to write the general solution for $\delta_{\rm dm}$ in terms of hypergeometric series, provided some conditions on DE equation of state $w$ are satisfied. These are of course general conditions on the parameters $\alpha$, $\beta$ e $\gamma$ which are described in \cite{abramowitz1964handbook}:
\begin{enumerate}
 \item If $\gamma$ is not an integer, i.e.
\begin{equation}
 w \neq \frac{1}{3(2n - 1)}\;, \qquad n = 0, \pm 1, \pm 2, ...\;,
\end{equation}
since $\gamma -\alpha - \beta = 1/2$, the hypergeometric series is well-defined and convergent for $y < 1$,  and we can write the general solution of \eqref{eqsubhordmnode3} as
\begin{eqnarray}\label{eqsubhordmnode3solbis}
 \delta_{\rm dm} = c_1\;F\left(-\frac{1}{3w},\frac{1}{2w};\frac{1}{2} + \frac{1}{6w};-y\right)\nonumber\\
 + c_2\;y^{\frac{1}{2} - \frac{1}{6w}}F\left(\frac{1}{2} - \frac{1}{2w},\frac{1}{2} + \frac{1}{3w};\frac{3}{2} - \frac{1}{6w}; -y\right)\;,
\end{eqnarray}
for $y < 1$, i.e. in the remote future. Note that $c_1$ and $c_2$ are integration constants. By means of Kummer transformations, one can also write the above hypergeometric series about infinity (this case, however, requires the additional condition that $\alpha - \beta = -1/6w$ is not an integer), i.e. in the regime where CDM dominates.

\item If it happens that one of $\alpha$, $\beta$, $\gamma - \alpha$ or $\gamma - \beta$ is an integer, then the above solutions \eqref{eqsubhordmnode3solbis} and its Kummer transformations still hold, only that they have a simpler form because one of the hypergeometric series would be truncated to a polynomial.

\item The case $\gamma = 1$ implies $w = 1/3$, i.e. the M\'esz\'aros case, which under the viewpoint of this classification is quite special. If we used solution eq.~\eqref{eqsubhordmnode3solbis} we would lose one of the two independent solutions. Indeed, we would just be left with
\begin{equation}
 \delta_{\rm dm} = c_1F(-1,3/2,1,-y) = c_1\left(1 + \frac{3}{2}y\right)\;,
\end{equation}
which is the growing mode. The decaying one can be found as
\begin{equation}
 \delta_{\rm dm} = c_2\left(1 + \frac{3}{2}y\right)\ln\left[\frac{\sqrt{1 + y} + 1}{\sqrt{1 + y} - 1}\right] - 3\sqrt{1 + y}\;.
\end{equation}

\item When $\gamma \ge 2$ is an integer, then
\begin{equation}
 w = \frac{1}{3(2n - 1)} = \frac{1}{9}, \frac{1}{15}, \frac{1}{21}, ...\;, \qquad n = 2, 3, 4...\;,
\end{equation}
and
\begin{equation}
 \alpha = 1 - 2n\;, \qquad \beta = \frac{3(2n - 1)}{2}\;, \qquad n = 2, 3, 4...\;,
\end{equation}
therefore $\alpha$ is a negative integer. An independent solution still is
\begin{equation}
 \delta_{\rm dm} = c_1F\left(1 - 2n, \frac{3(2n - 1)}{2}, n, -y\right)\;,
\end{equation}
which is therefore a polynomial, whereas the other independent solution has a complicated form given in terms of $\psi$ (digamma) functions. For more detail, we refer the reader to \cite{abramowitz1964handbook}.

\item When $\gamma \le 0$ is an integer, then
\begin{equation}
 w = -\frac{1}{3(2m + 1)} = -\frac{1}{3}, -\frac{1}{9}, -\frac{1}{15}, ...\;, \qquad m = 0, 1, 2...\;,
\end{equation}
and
\begin{equation}
 \alpha = 1 + 2m\;, \qquad \beta = - \frac{3(2m + 1)}{2}\;, \qquad m = 0, 1, 2...\;.
\end{equation}
Since $\gamma$ is a negative integer and none between $\alpha$ and $\beta$ is also a negative integer, the hypergeometric series in eq.~\eqref{eqsubhordmnode3solbis} are not defined. An independent solution is given by
\begin{equation}
 \delta_{\rm dm} = c_1y^{m + 1}F\left(2 + 3m, -\frac{4m + 1}{2}, 2 + m, -y\right)\;,
\end{equation}
whereas the other has again a complicated form given in terms of $\psi$ functions, for which we refer the reader to \cite{abramowitz1964handbook}.
\end{enumerate}
We have thus exhausted the classification of the possible solutions of the $w$-M\'esz\'aros equation. Being our interest in a dark component whose $w$ is about $-1$, we focus our investigation on solution~\eqref{eqsubhordmnode3solbis}.
\section{Asymptotic behaviour and matching conditions}\label{sec:5}

Let us consider the first terms in the hypergeometric series of eq.~\eqref{eqsubhordmnode3solbis}:
\begin{eqnarray}
 F\left(-\frac{1}{3w},\frac{1}{2w};\frac{1}{2} + \frac{1}{6w};-y\right) = 1 + \frac{y}{w(1 + 3w)} + O(y^2)\;,\\
y^{\frac{1}{2} - \frac{1}{6w}}F\left(\frac{1}{2} - \frac{1}{2w},\frac{1}{2} + \frac{1}{3w};\frac{3}{2} - \frac{1}{6w};-y\right) = y^{\frac{1}{2} - \frac{1}{6w}}\left[1 + O(y)\right]\;.
\end{eqnarray}
When $y \to 0$, i.e. in the pure dark energy dominated epoch, it appears that $\delta_{\rm dm} \to c_1$, the perturbation in the matter component tends to a constant value. The other solution scales as
\begin{equation}
 \delta_{\rm dm} = c_2\;y^{\frac{1}{2} - \frac{1}{6w}}\;, \qquad (y \ll 1)\;,
\end{equation}
or, in the scale factor
\begin{equation}
 \delta_{\rm dm} = c_2 \left(\frac{a}{a_{\rm de}}\right)^{\frac{3w}{2} - \frac{1}{2}}\;, \qquad (a \gg a_{\rm de})\;,
\end{equation}
where remember that $w < -1/3$, so it is a decaying mode. However, some care must be taken when considering the limit $y \to 0$ because each perturbation mode is destined to exit the Hubble horizon, where our approximation of sec.~3 no longer holds true. This can be seen by using Friedmann equation \eqref{FriedmEq} and writing explicitly the $a$-dependence of the Hubble horizon:
\begin{equation}
\frac{1}{H} \sim a^{\frac{3}{2}(1 + w)}\;, \qquad (a \gg a_{\rm de})\;.
\end{equation}
Since $w < -1/3$, the above exponent is less than unity, whereas any scale grows proportionally to $a$ and therefore shall exit the horizon in due time. We shall consider our results only up to $a = 1$.

We now investigate the $y \to \infty$ behaviour in order to match solution \eqref{eqsubhordmnode3mattersol} with eq.~\eqref{eqsubhordmnode3solbis}, which we rewrite as 
\begin{eqnarray}\label{eqsubhordmnode3solter}
 \delta_{\rm dm} = c_1D_1(y) + c_2D_2(y)\;,
\end{eqnarray}
i.e. with the identification:
\begin{equation}
 D_1(y) \equiv F\left(-\frac{1}{3w},\frac{1}{2w};\frac{1}{2} + \frac{1}{6w};-y\right)\;,
\end{equation}
and
\begin{equation}
 D_2(y) \equiv y^{\frac{1}{2} - \frac{1}{6w}}F\left(\frac{1}{2} - \frac{1}{2w},\frac{1}{2} + \frac{1}{3w};\frac{3}{2} - \frac{1}{6w}; -y\right)\;.
\end{equation}
Our matching conditions require that for a given $y_m \gg 1$, we should have:
\begin{equation}
 A\;y_m^{-1/2w} + B\;y_m^{1/3w} = c_1D_1(y_m) + c_2D_2(y_m)\;.
\end{equation}
We do not need to match the derivatives, since the asymptotic behaviour is polynomial. The asymptotic behaviour of the hypergeometric functions is:
\begin{eqnarray}
 D_1(y_m) \sim G_1\;y_m^{-1/2w} + G_2\;y_m^{1/3w}\;,\\
 D_2(y_m) \sim G_3\;y_m^{-1/2w} + G_4\;y_m^{1/3w}\;,
 \end{eqnarray}
where:
\begin{eqnarray}
 G_1 := \frac{\Gamma(1/2 + 1/6w)\Gamma(-5/6w)}{\Gamma(1/2-1/3w)\Gamma(-1/3w)}\;, \qquad
 G_2 := \frac{\Gamma(1/2 + 1/6w)\Gamma(5/6w)}{\Gamma(1/2 + 1/2w)\Gamma(1/2w)}\;,\\
 G_3 := \frac{\Gamma(3/2 - 1/6w)\Gamma(-5/6w)}{\Gamma(1/2-1/2w)\Gamma(1-1/2w)}\;, \qquad
 G_4 := \frac{\Gamma(3/2 - 1/6w)\Gamma(5/6w)}{\Gamma(1/2 + 1/3w)\Gamma(1 + 1/3w)}\;,
\end{eqnarray}
where $\Gamma$ is Euler's gamma function. We thus find:
\begin{equation}
 A = c_1G_1 + c_2G_3\;, \qquad B = c_1G_2 + c_2G_4\;,
\end{equation}
from which
\begin{equation}\label{junccondc1c2}
 c_1 = \frac{AG_4 - BG_3}{G_1G_4 - G_2G_3}\;, \qquad c_2 = \frac{BG_1 - AG_2}{G_1G_4 - G_2G_3}\;.
\end{equation}
The coefficient $A$ and $B$ carry a functional dependence on $k$ which is an inheritance of the radiation-dominated phase, where the evolution of $\delta_{\rm dm}$ indeed depended on $k$, see \cite{Weinberg:2002kg}. On the other hand, during the matter era this functional dependence is not modified on any scales. Moreover, the same functional dependence is also not modified during the dark energy era, provided we stay on small scales and $c_{\rm de}^2$ is not too small. In this situation, the transfer function $T(k)$ is not modified as a function of $k$, but only via the growth function which depends on $w$.

\section{Goodness of the approximated solution, growth function, logarithmic growth function and growth index parameter}\label{sec:gamma}

We focus on the approximated solution for $A = 0$, i.e. neglecting the decaying mode of the matter-dominated era. We have from eq.~\eqref{junccondc1c2}:
\begin{equation}
 c_1 = \frac{-BG_3}{G_1G_4 - G_2G_3}\;, \qquad c_2 = \frac{BG_1}{G_1G_4 - G_2G_3}\;,
\end{equation}
where $B$ is determined by the initial condition. These formulas can be simplified as:
\begin{eqnarray}
\label{c1form} c_1 = B\frac{\sqrt{\pi } 2^{\frac{w-1}{w}} \csc \left(2\pi/3w\right) \Gamma \left(1/w\right)}{\left[\csc\left(\pi/w\right)+\csc \left(2 \pi/3 w\right)\right] \Gamma \left[\left(3+1/w\right)/6\right] \Gamma \left(5/6 w\right)}\;,\\
\label{c2form} c_2 = B\frac{\sqrt{\pi } 2^{-\frac{2}{3 w}} \csc \left(\pi/w\right) \Gamma \left(1+ 2/3 w\right)}{\left[\csc \left(\pi/w\right)+\csc \left(2 \pi/3 w\right)\right] \Gamma \left(3/2-1/6 w\right) \Gamma \left(5/6 w\right)}\;.
\end{eqnarray}
Substituting these in eq.~\eqref{eqsubhordmnode3solter}, we obtain the main result of this paper, i.e. an analytic formula for the DM density contrast:
\begin{eqnarray}\label{anformgrowth}
\delta_{\rm dm}^{*}(w,y) = c_1F\left(-\frac{1}{3w},\frac{1}{2w};\frac{1}{2} + \frac{1}{6w};-y\right) +\nonumber\\ c_2y^{\frac{1}{2} - \frac{1}{6w}}F\left(\frac{1}{2} - \frac{1}{2w},\frac{1}{2} + \frac{1}{3w};\frac{3}{2} - \frac{1}{6w}; -y\right)\;,
\end{eqnarray}
where the star serves to distinguish it from the numerical solution, with which we will compare it in order to assess the goodness of this approximation. In the case $w = -1$, i.e. the $\Lambda$CDM model, the above approximation becomes
\begin{equation}
 \delta_{\rm dm}^{*}(w = -1, y) = \frac{3B\Gamma\left(2/3\right)^2}{2^{2/3}\Gamma\left(-5/3\right)}\sqrt{1 + y} - \frac{5B}{4}y^{2/3}F\left(1/6,1;5/3;-y\right)\;,
\end{equation}
where $y = \Omega_{\rm dm,0}a^{-3}/\left(1 - \Omega_{\rm dm,0}\right)$, for $w = -1$.

\begin{figure}[htbp]
\center
\includegraphics[width=0.48\textwidth]{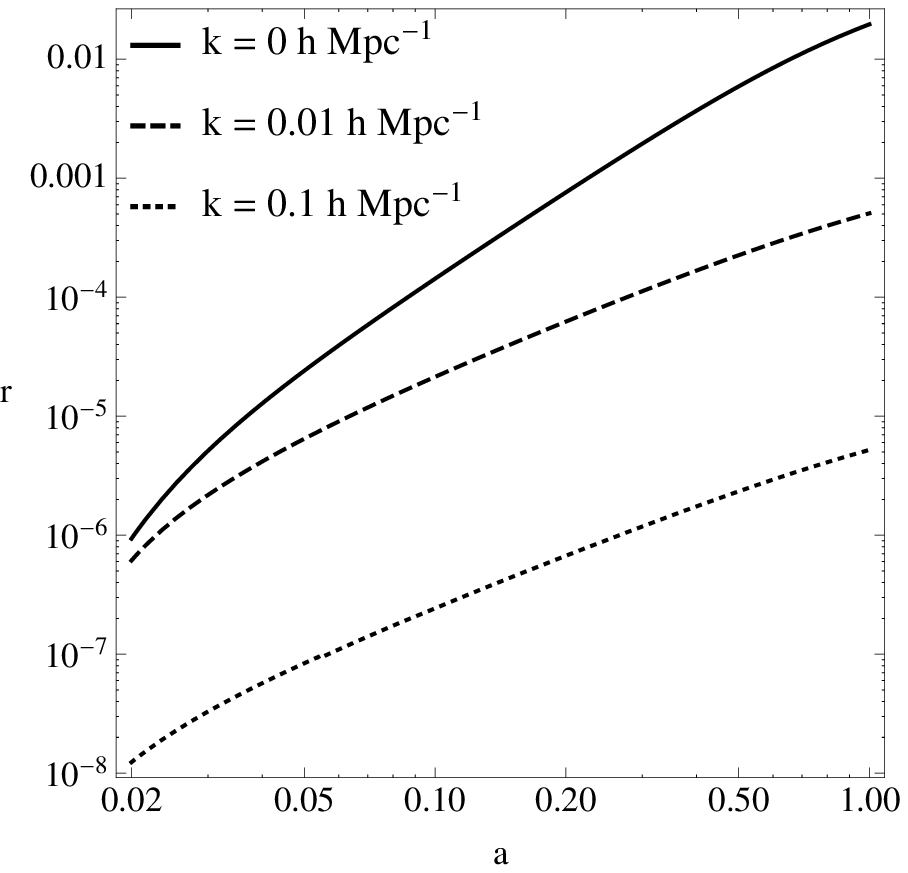}\includegraphics[width=0.485\textwidth]{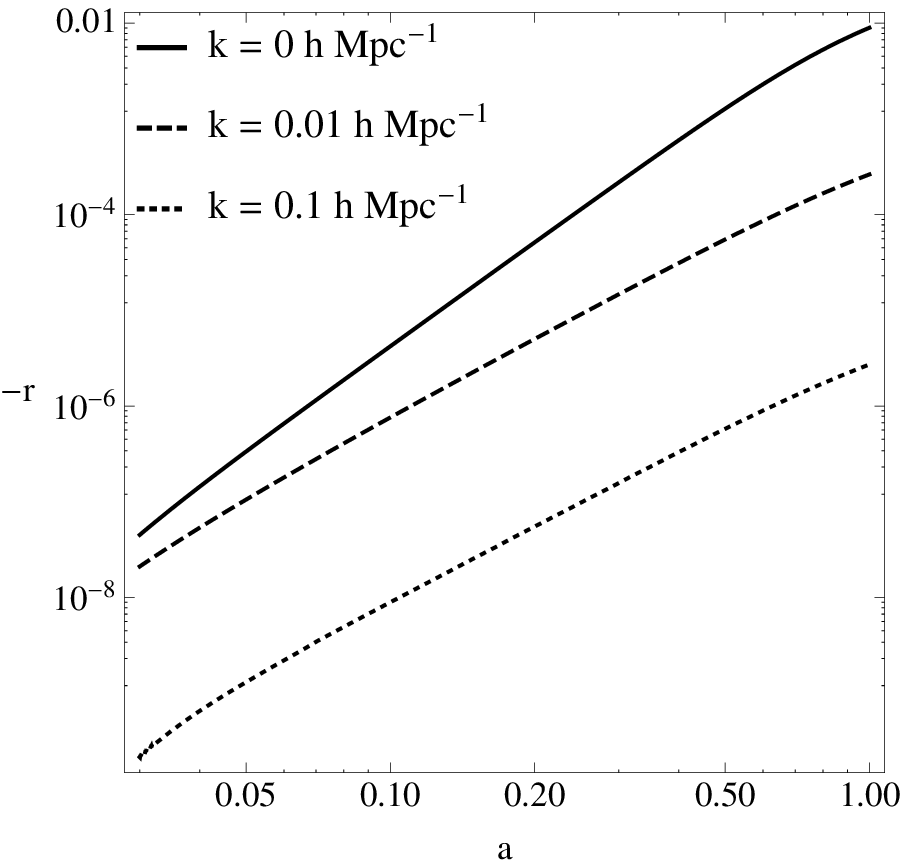}\\
\includegraphics[width=0.48\textwidth]{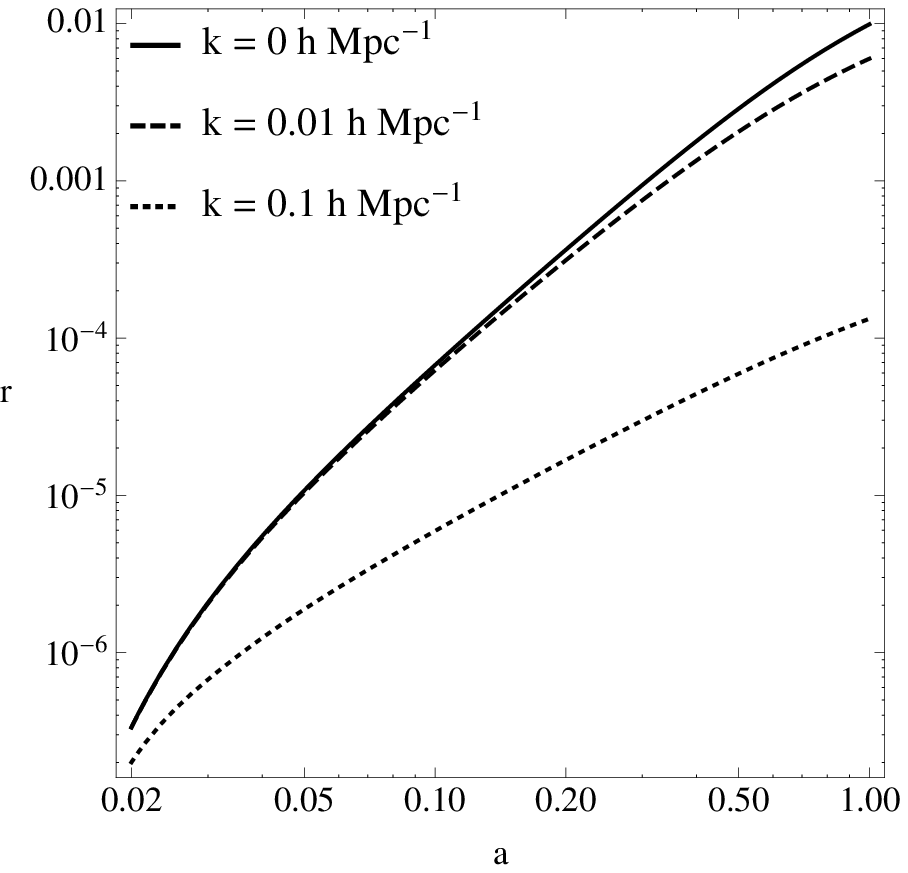}\includegraphics[width=0.495\textwidth]{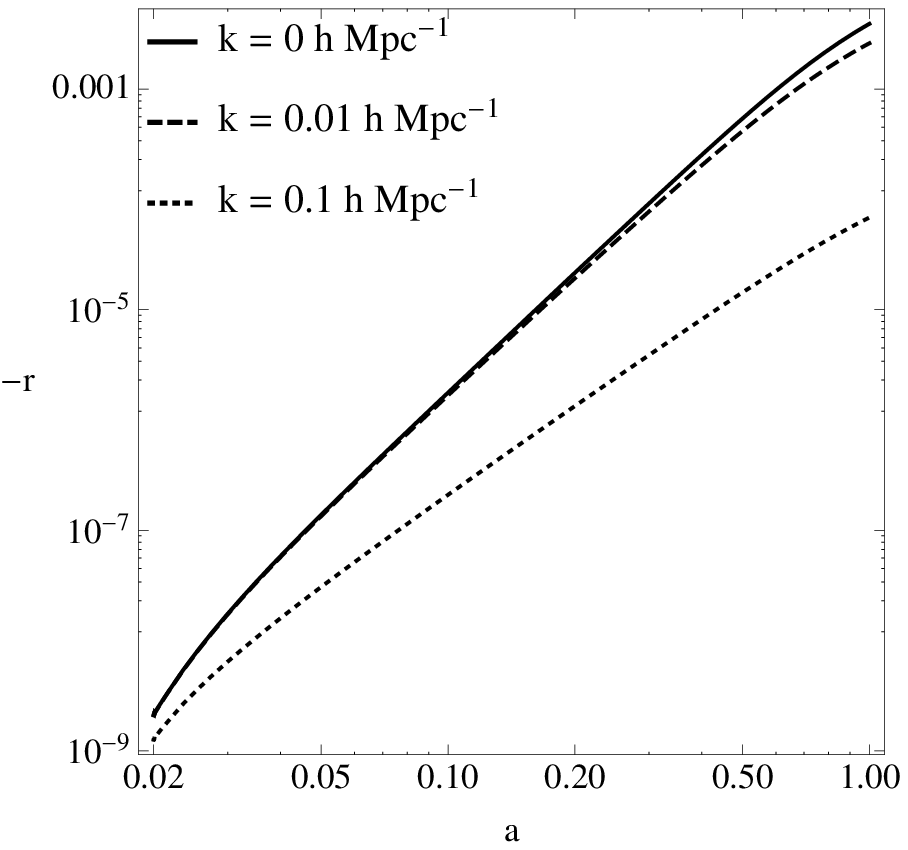}
\caption{Evolution of $r$ as a function of $a$  and for $k = 0, 0.01, 0.1$ h Mpc$^{-1}$ (solid line, dashed line and dotted line, respectively) for $w = -0.8$ (left panels) and $w = -1.2$ (right panels). The equivalence scale factor $a_{\rm de}$ is computed from eq.~\eqref{OmegaD0aeqrel} for the fiducial model $\Omega_{\rm dm,0} = 0.3$, i.e. $a_{\rm de} = 0.70$ for $w = -0.8$ and $a_{\rm de} = 0.79$ for $w = -1.2$. The DE effective speed of sound chosen here is $c_{\rm de}^2 = 1$ (upper panels) and $c_{\rm de}^2 = 0.01$ (lower panels).}
\label{figComp}
\end{figure}

In order to compute the numerical solution for the DM density contrast, we rewrite \eqref{Eq1}-\eqref{Eq4} in the following form:
\begin{eqnarray}
\label{Eq1bis} \delta_{\rm dm,aa} + \left(\frac{H_a}{H} + \frac{3}{a}\right)\delta_{\rm dm,a} = \frac{3H_0^2}{2H^2a^2}\left[\Omega_{\rm dm}\delta_{\rm dm} + \left(1 + 3c_{\rm de}^2\right)\Omega_{\rm de}\delta_{\rm de}\right]\;,\\
\label{Eq2bis} \delta_{\rm de,aa} + \left[\frac{3}{a}\left(1 + c_{\rm de}^2 - 2w\right) + \frac{H_a}{H}\right]\delta_{\rm de,a} + \frac{3}{a}\left(c_{\rm de}^2 - w\right)\left(\frac{2}{a} - \frac{3w}{a} + \frac{H_a}{H}\right)\delta_{\rm de} = \nonumber\\ \frac{3H_0^2(1 + w)}{2H^2a^2}\left[\Omega_{\rm dm}\delta_{\rm dm} + \left(1 + 3c_{\rm de}^2\right)\Omega_{\rm de}\delta_{\rm de}\right] - \frac{3w(1 + w)}{a}\delta_{\rm dm,a} - \frac{k^2c_{\rm de}^2}{H^2a^4}\delta_{\rm de}\;,
\end{eqnarray}
where a subscript $a$ means derivation with respect to the scale factor. We solve this numerical system starting from a scale factor $a_i = 0.01$, using initial conditions $\delta_{\rm de}(a_i) = 0$, $\delta_{\rm de,a}(a_i) = 0$ for the DE density contrast and for the DM density contrast we use the same values computed from \eqref{anformgrowth} for $a_i = 0.01$ and for $B = 1$. We define the following quantity:
\begin{equation}
r \equiv \frac{\delta_{\rm dm} - \delta_{\rm dm}^{*}}{\delta_{\rm dm}}\;,
\end{equation}
as an indicator of the goodness of the approximation and plot its values in figure \ref{figComp} for $c_{\rm de}^2 = 1$ and for $c_{\rm de}^2 = 0.01$.
It is impressive to notice that even in the case $k = 0$, which is totally out of our hypothesis of sub-horizon perturbations, the discrepancy between \eqref{anformgrowth} and the numerical solution is less than 1\%. Notice that in the right panels of figure \ref{figComp} we have plotted $-r$ because our approximation overestimates the numerical solution. It is also interesting to notice that for $c_{\rm de}^2 = 0.01$ the agreement between numerical solution and analytic approximation is worse than in the $c_{\rm de}^2 = 1$ case, as expected since DE perturbations act efficiently on smaller scales.

\subsection{Growth function, logarithmic growth function and growth index}

\begin{figure}[htbp]
\center
\includegraphics[width=0.48\textwidth]{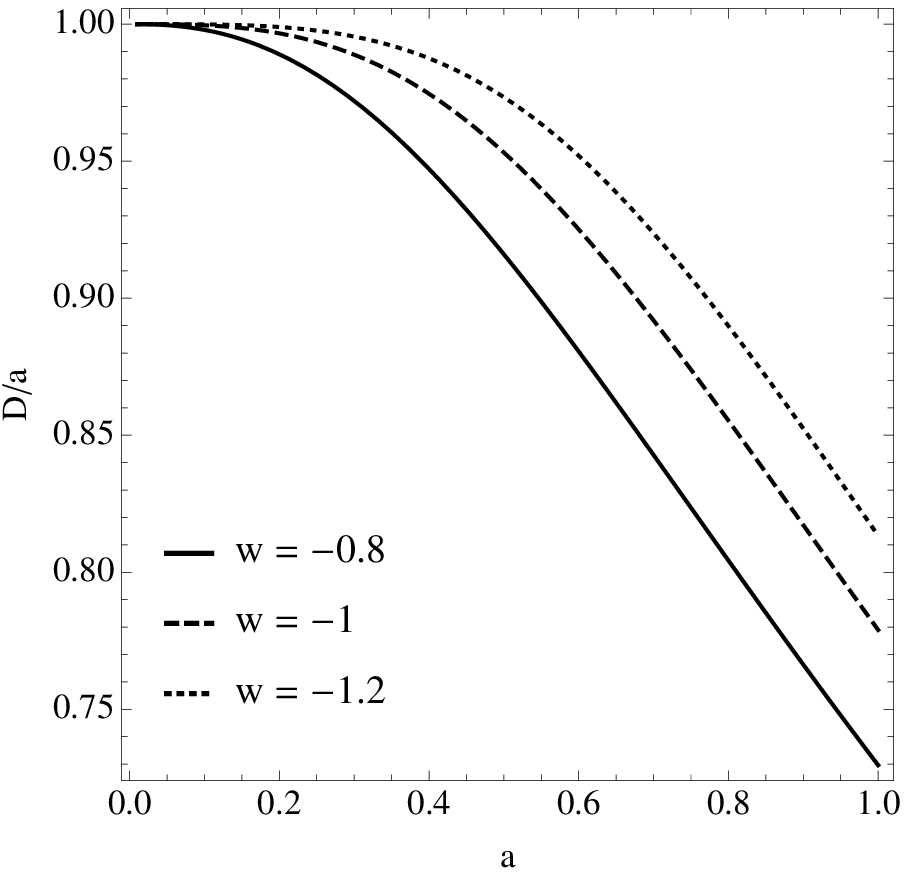}\includegraphics[width=0.485\textwidth]{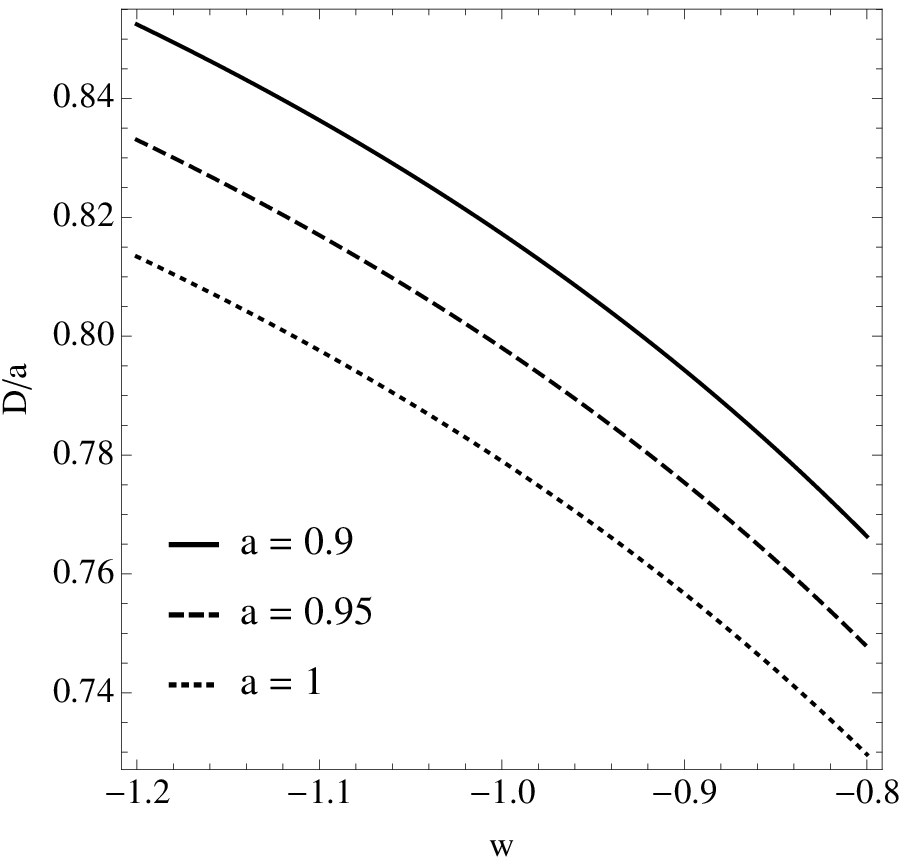}\\
\caption{Left panel: Evolution of the growth factor $D$, normalized to $a$, as a function of $a$ for $w = -0.8$ (solid line), $w = -1$ (dashed line) and $w=-1.2$ (dotted line). Right panel: Evolution of $D$ normalized to $a$ as a function of $w$ for $a = 0.9$ (solid line), $a = 0.95$ (dashed line) and $a = 1$ (dotted line). The equivalence scale factor $a_{\rm de}$ is computed from eq.~\eqref{OmegaD0aeqrel} for the fiducial model $\Omega_{\rm dm,0} = 0.3$, i.e. $a_{\rm de} = 0.70$ for $w = -0.8$, $a_{\rm de} = 0.75$ for $w = -1$ and $a_{\rm de} = 0.79$ for $w = -1.2$. }
\label{fig1}
\end{figure}

If we choose $B$ in order to normalize the initial value of $\delta_{\rm dm}^*$ to unity, we obtain what is generally known as growth function, denoted with $D$. We plot it in the left panel of figure \ref{fig1}, normalized to the linear growth, for reference. In the right panel of the same figure we plot $D$ as a function of $w$ for different values of $a$. As before, we choose $a_i = 0.01$.
Note some features in these plots. First of all, when $w < - 1$, the growth is larger than in the case $w > -1$. This happens because DE is less dominant in the past and thus affects the growth of dark matter inhomogeneities in a weaker way. Second, increasing the equivalence scale factor diminishes the decay of the growth factor. This is easily understandable, since the more recent the equality is, the less dark energy has dominated and thus had time to thwart the growth of matter inhomogeneities. The growth function $D/a$ depends appreciably from $w$. Indeed for the wide range of values chosen, i.e. $-1.2 < w < -0.8$, $D/a$ varies of about 20\%. Note that the results given in the Introduction for $w_0$ (which is equal to our $w$ since the latter is constant) have also uncertainties of about 20\% (at 95\% CL).

\begin{figure}[h]
\includegraphics[width=0.48\textwidth]{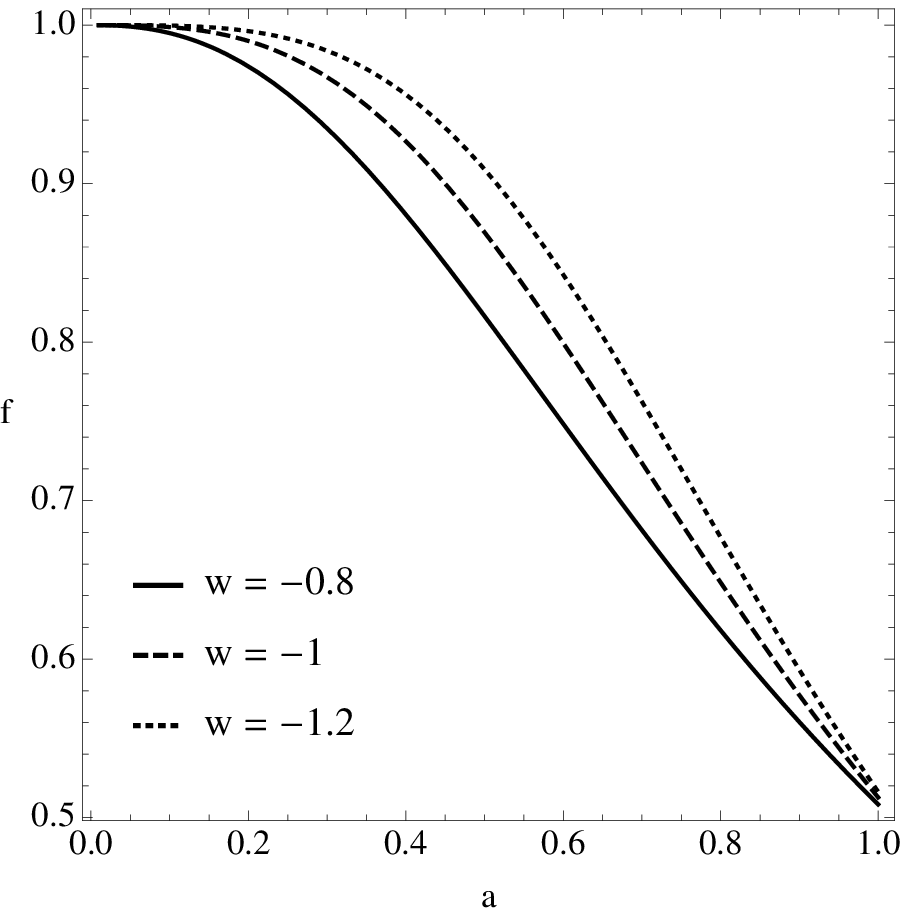}\includegraphics[width=0.505\textwidth]{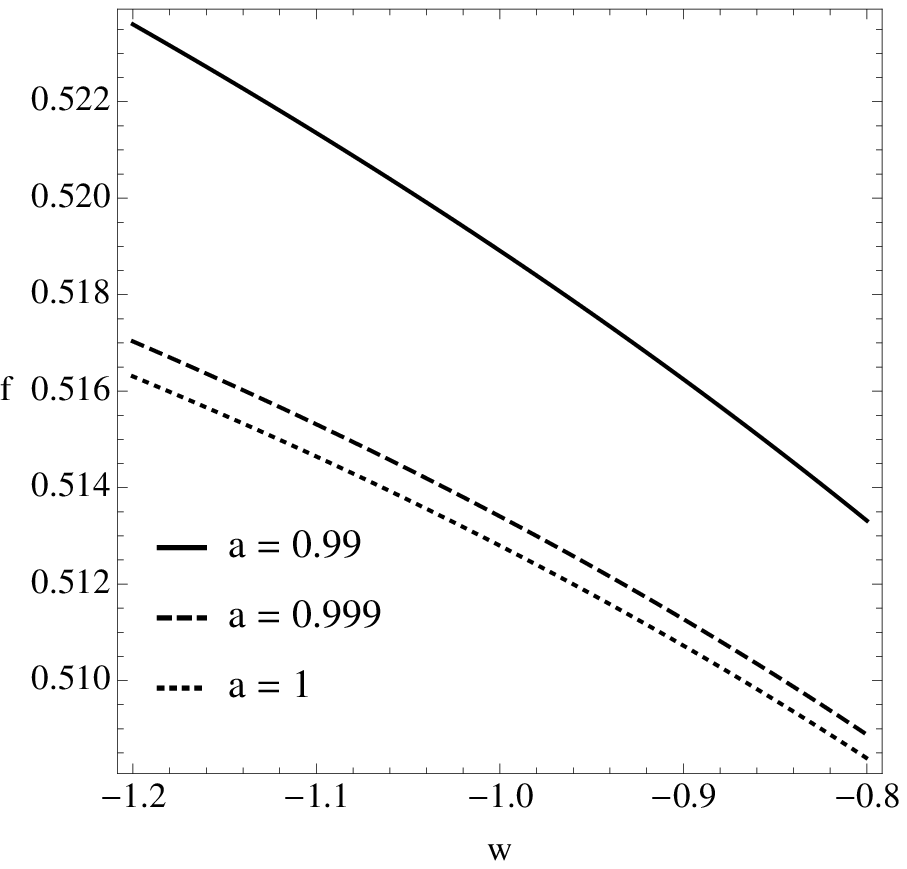}
\caption{Left panel: Evolution of $f$ as function of $a$ for $w = -0.8$ (solid line), $w = -1$ (dashed line) and $w =-1.2$ (dotted line). Right panel: Evolution of $f$ as a function of $w$ for $a=0.99$ (solid line), $a = 0.999$ (dashed line) and $a = 1$ (dotted line). The equivalence scale factor $a_{\rm de}$ is computed from eq.~\eqref{OmegaD0aeqrel} for the fiducial model $\Omega_{\rm dm,0} = 0.3$.}
\label{figf}
\end{figure}

\begin{figure}[htbp]
\includegraphics[width=0.48\textwidth]{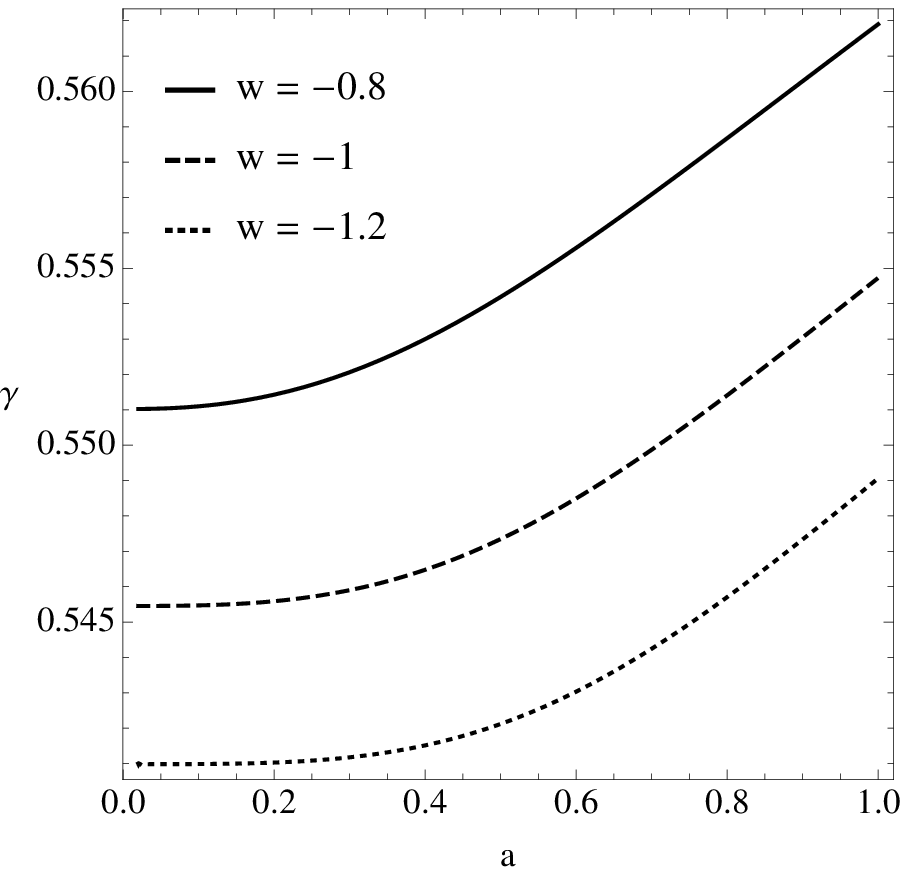}\includegraphics[width=0.485\textwidth]{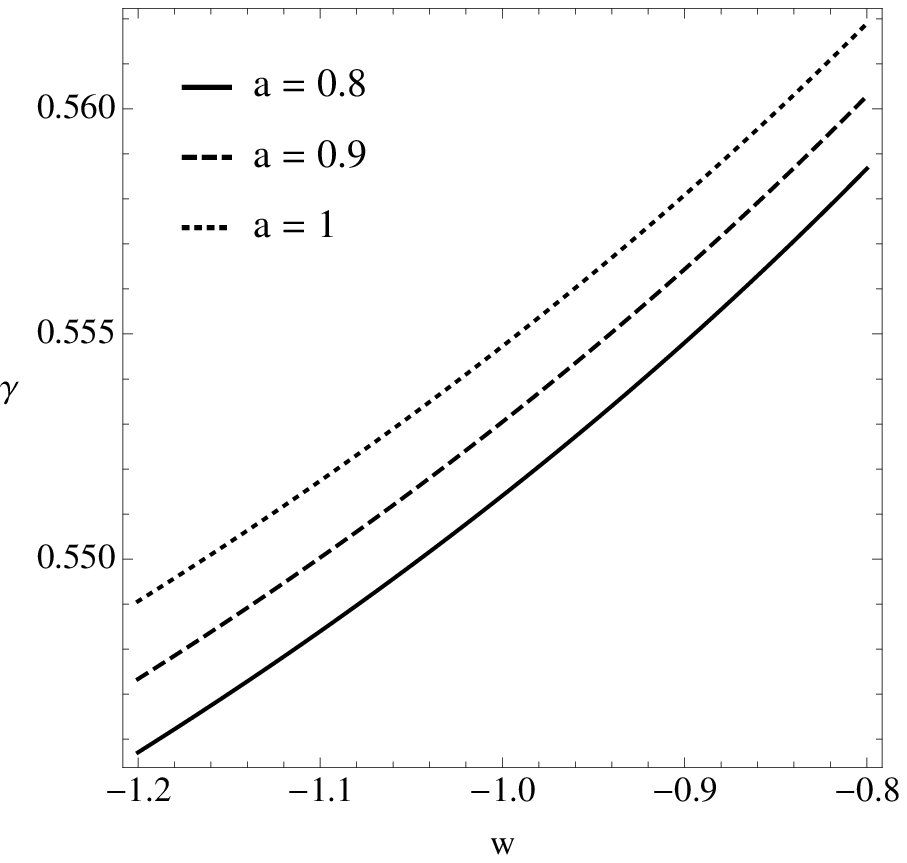}
\caption{Left panel: Evolution of $\gamma$ as function of $a$ for $w = -0.8$ (solid line), $w = -1$ (dashed line) and $w =-1.2$ (dotted line). Right panel: Evolution of $\gamma$ as a function of $w$ for $a = 0.8$ (solid line), $a = 0.9$ (dashed line) and $a = 1$ (dotted line). The equivalence scale factor $a_{\rm de}$ is computed from eq.~\eqref{OmegaD0aeqrel} for the fiducial model $\Omega_{\rm dm,0} = 0.3$.}
\label{figgamma}
\end{figure}

It is also interesting to plot the form of the logarithmic growth function and the growth index function, defined as:
\begin{equation}
f = \frac{d\ln\delta_{\rm dm}}{d\ln a} = \frac{d\ln D}{d\ln a}\;, \qquad \gamma = \frac{\ln f}{\ln \tilde\Omega_{\rm dm}}\;,
\end{equation}
where
\begin{equation}
\tilde\Omega_{\rm dm} = \frac{\rho_{\rm dm}}{\rho_{\rm dm} + \rho_{\rm de}} = \left(1 + \frac{1 - \Omega_{\rm dm,0}}{\Omega_{\rm dm,0}}a^{-3w}\right)^{-1}\;.
\end{equation}
In fig.~\ref{figf} we plot $f$ and in fig.~\ref{figgamma} we plot $\gamma$. The results found here from our analytic formula are in very good agreement with those that can be found in the literature, see for example \cite{Dossett:2013npa, Batista:2014uoa}.

In principle, from the analytic formula \eqref{anformgrowth} it is possible to derive also analytic formulas for $f$ and $\gamma$, but in practice they are so cumbersome that perhaps cannot be very useful. On the other hand, the usefulness of an analytic formula stays also in the possibility of finding a series expansion. For example, the series expansion for $f$ for small values of $y$ is the following: 
\begin{equation}
f = \frac{3y}{1 + 3w} + O\left(y^2\right) + \left[\frac{2}{-1 + 3w}\left(1 + \frac{c_1}{c_2}y^{1/6w - 1/2}\right) + O\left(y^{1/6w + 1/2}\right)\right]^{-1}\;,
\end{equation}
where from eqs.~\eqref{c1form} and \eqref{c2form}
\begin{equation}
\frac{c_1}{c_2} = -\frac{2^{1-1/3w}\Gamma\left(3/2 - 1/6w\right)\Gamma\left(-2/3w\right)}{\Gamma\left(1/2 + 1/6w\right)\Gamma \left(1-1/w\right)}\;.
\end{equation}
Note that since $y_0 = \Omega_{\rm dm,0}/\left(1 - \Omega_{\rm dm,0}\right)$, i.e. $y_0 = 3/7$, for the fiducial model $\Omega_{\rm dm,0} = 0.3$, the truncation error in the above formula may be quite large, e.g. 75\% for $w = -1$.

%%%%%%%%%%%%%%%%%%%%%%%%%%%%%%%%%%%%%%%%%%%%%%%%%%%%%%%%%%%%%%%%%%%%%%%%%%%%%%%%%%%%%%%%%%%%%%%%%%%%%%%%%%%%%%%%

\section{Discussion and conclusion}\label{sec:6}

We investigated the evolution of density fluctuations in cold dark matter through the epoch of equivalence between dark matter and dark energy. We assumed, for dark energy, a constant equation of state $w$ and a constant effective speed of sound $c_{\rm de}^2$. In order to perform analytic calculations, we considered perturbations well inside the Hubble horizon ($k/a \gg H$) and, in order to neglect dark energy fluctuations, we assumed $c_{\rm de}^2$ to be of order unity. Working on the evolution equations for perturbations, we obtained an equation for the density contrast of cold dark matter, called $w$-Mészáros equation, eq.~\eqref{eqsubhordmnode3}. It can be cast in the form of a Gaussian hypergeometric equation and thus solved analytically. We classified the solutions depending on the value of $w$, and chose the relevant one, in agreement with observation. By matching the solution in the matter dominated era, we then calculated exactly the growth function, logarithmic growth function and growth index for cold dark matter in presence of dark energy, eq.~\eqref{anformgrowth}. We then assess the goodness of our analytic approximation and find an excellent agreement with the numerical solution, being the discrepancy less than 1\% even for the case $k = 0$, which corresponds to super-horizon scales, i.e. out of our approximation.

%%%%%%%%%%%%%%%%%%%%%%%%%%%%%%%%%%%%%%%%%%%%%%%%%%%%%%%%%%%%%%%%%%%%%%%%%%%%%%%%%%%%%%%%%%%%%%%%%%%%%%%%%%%%%%%%

\section*{Acknowledgements}

The authors thank CNPq (Brazil) and Fapes (Brazil) for partial financial support. The authors are thankful to Daniele Bertacca, Hermano E. S. Velten and Júlio C. Fabris for useful comments and suggestions.

%%%%%%%%%%%%%%%%%%%%%%%%%%%%%%%%%%%%%%%%%%%%%%%%%%%%%%%%%%%%%%%%%%%%%%%%%%%%%%%%%%%%%%%%%%%%%%%%%%%%%%%%%%%%%%%%

\bibliographystyle{unsrt}

\end{document}